\documentstyle[12pt]{article}

\begin{document}
\newcommand{\be}{\begin{eqnarray}}
\newcommand{\ee}{\end{eqnarray}}
\begin{center}
   {\large \bf PHYSICS OF HEAVY IONS COLLISIONS: \\ THE SUMMARY OF MORIOND-97}\\[5mm]
   E.V. SHURYAK\\[5mm]
   {\small \it  Department of Physics\\
   SUNY at Stony Brook, Stony Brook, NY11790, USA \\[8mm] }
\end{center}

\begin{abstract}\noindent
   CERN dilepton experiments have provided the most exciting data. Strong enhancement at low masses observed by CERES and HELIOS3 indicate strong modification
in  the vector channel in matter compared to vacuum properties. NA50
data on $J/\psi$ suppression in PbPb collisions show surprising
deviation from the previous trend. The question is whether it is the
expected early-time signal of QGP, or due to late-time hadronic interactions.
 Theoretical  and experimental suggestions
have been made to resolve this   issue. BNL and SPS experiments have 
also provided rather complete data with heavy beams (Au and Pb, respectively).
Very strong collective  flow effects have been observed at both energies,
which allow for the first time to restrict the EOS of the hadronic  
matter.
Several observables (flow, Coulomb effects and HBT) suggest rather
long evolution of systems created in heavy ion collisions and very low
freeze-out densities relative to previous studies.
  Theory of jet stopping in QGP is becoming quantitative.

\end{abstract}
\newpage
\section{Introduction}
  At this meeting the heavy ion physics was represented much more than
  at any previous Morionds, but it is still a minority and therefore
it is probably necessary to start this general talk by reminding $why$ these
experiments are done. Many high energy physicists,  proud of
their accurate studies of hard processes  and (truly impressive!)
agreement with the perturbative QCD, are asking if one really needs  to
find a quark-gluon plasma (QGP). My answer to that is that studies of
  the QCD phase transition is only partly related to matter properties at
very high T. A very significant part of the  
motivation is a  desire to understand how and why the ordinary
hadrons ``melt'', and especially how  the
ground state (the ``QCD vacuum'') can be rearranged under these conditions.
 It is widely believed that it 
helps to understand hadronic and vacuum structure of QCD (and maybe
non-perturbative phenomena in gauge theories in general). To my mind
these topics 
 are the very core of the scientific challenge which
makes QCD  interesting: but (very
characteristically for these days)
it was never even touched in the high energy part of the meeting.
  
To explain what I mean,
let me use a very elementary example.
The pressure of
the QGP is $p=O(T^4)-B$, where
the first (simple) term represent
perturbative quarks and gluons, while the second (non-trivial)
 bag term\footnote{One should not confuse the value of this
B with the much smaller one found in the fit to the MIT bag model.} represents a price for suppressing
non-perturbative vacuum fluctuations in QGP.
 Now recall textbook ``vacuum engineering"
experiment with Magdeburg semi-spheres: after one pumps the air out of
them, one need a very large force to separate them. 
Indeed,
in the equation above the first quark-gluon thermal
 term is simply the analog of horses,
 working against the vacuum pressure and making expansion possible.
Their strong compensation (or ``softness" of EOS in the
phase transition region) is what we seem to observe now. 

 At this time  it  is
really a pleasure to summarize the heavy ion part of the conference:
there are
 so many excited new observations. 
 Very different explanations of what they mean have
been put forward, making discussion extremely  interesting.
Clearly it was a very busy meeting, and
   too many interesting  talks I am not able even to mention, keeping
to few main subjects. Fortunately,    some
  topics had
been  summarized elsewhere. For example, theory of
the two-particle interferometry (HBT) was
nicely covered in  \cite{HBTt} and experiment
in \cite{Foka,Gaardhoje}
. Recent progress in
understanding of the microscopic mechanism of the chiral
 phase transition was covered in my original
talk, and I would not comment on it here.
    
\section{Dileptons and photons}
 
  Already at QM93, the HELIOS3 experiment has reported very interesting data,
and in \cite{SX_where} it was emphasized that those significantly exceed  
   expectations from ``basic conventional sources'',
the pion annihilation in 
the hadronic phase and $\bar q q\rightarrow e^+e^-$ in QGP. 
 Later CERES experiment  (for current status see \cite{Pfeiffer})
  has observed even more dramatic excess of dileptons at
  $M_{e+e-}\approx 400-600 MeV $. Since dileptons are ``penetrating
probes", the excess production  happens early in the collision, and
the issue of rescattering ( which is so painful to settle for $J/\psi$)
does not arise here.

  Multiple attempts by theorists to explain it were discussed
in some details at this meeting \cite{Redlich}.
In short, all attempts to use other ``conventional sources'' has basically failed
\footnote{In fact, I have never before  seen
  that rather involved calculations by several groups agree so well,
  not only in the shape of spectra of $M_{e+e-}$  but also in absolute normalization. }.
  A list of ``unconventional'' explanations 
include: (i) dropping $m_\rho$ due to chiral phase transition
   \cite{LKB_95}; (ii) increasing width of $\rho$
due to interaction with baryons \cite{Rapp} 
;
 (iii)
pion occupation numbers at low momenta are very high \cite{KKP_93}; 
(iv) 
 dropping  $m_{\eta'}$ \cite{KKM_96}. Let me comment on them subsequently.

  Li-Ko-Brown  has developed the well known idea of ``dropping rho mass"
into a detailed cascade model, explaining in details both all CERES and HELIOS3
data with one set of parameters. Many people have also tried to check this 
idea,
(including both  Redlich and myself \cite{HS_96} ), with the conclusion that
this idea  may indeed explain the observed mass spectrum.
In order to test it further, one may
 increase the resolution of CERES and check
how many $\rho$ remains in the peak, on the top of  the un-shifted 
 $
\omega$. A less trivial test \cite{HS_96}:
 if $m_\rho(T)$ is dropping, 
 that  of its axial partner
 $m_{A_1}$ should follow\footnote{Strict relation between the two were made
  e.g. in the contexts of Weinberg-type sum rules
   \cite{KS}. Those demand that both states  become
identical at $T_c$.}. As shown in  \cite{HS_96}, $a_1\rightarrow \pi e^+e^-$
decay populate low mass region and may become comparable or exceeding the background, if this mass is decreasing enough. 

 The explanation (ii)  \cite{Rapp} does not relate the low-mass enhancement with chiral
restoration, but  simply with the non-zero density of 
baryons.  As also commented by Redlich, this idea has not yet developed into
quite realistic model, and many questions about space-time evolution and
baryon composition remains. Nevertheless, their
 estimates suggest that at relevant
   conditions the rho width is 
   becoming large, even comparable to its mass. If so, one may argue that
$\rho$ simply ``melts", it  is no longer a well-defined resonance and its relevance may therefore be questioned\footnote{Significant vertex corrections due to modified rho,
 discussed by Redlich, is related to the same problem.}. However if
 one  uses simple
 ``duality" between very wide rho and $\bar q q $ continuum, it is
impossible to explain the data. What is needed here
is some new {\it dynamical correlation} in the vector channel,
which is absent in vacuum but appears in matter.
 The {\it instanton-antiinstanton
molecules}
(discussed in my talk) provide exactly that, a new attractive interaction
 close to the chiral restoration transition. However
whether it is sufficiently strong
to explain these data remains to be seen. 

  In any case,  one should test experimentally whether
the dilepton excess is indeed proportional to the baryon density, either 
by running at low energy at CERN,
or by waiting till HADES data.

   (iii) The observed low-$p_t$ excess of pions (over thermal
spectra) can indeed  dramatically increase the low-mass dilepton
yield. However (see my original talk), development of the non-zero
$pion$ chemical potential is the late-time effect,   
 which may hardly affect the early-stage dilepton 
production. 

(iv) It seems indeed very possible that
 $m_{\eta'}(T)$ should drop  more
than any other mass. But  in order to account for
 CERES excess
one should increase  yield of ``escaping  
 $\eta'$'' by  too huge a factor. It
contradicts to observed $\eta$ yield, so this idea does not work.

  The data from photon experiment WA98 
\cite{Peitzmann} on  $\pi^0$ spectrum of transverse momenta is now
reaching above $p_t=3 GeV$. It is probably  the first
data set  to enter the ``hard
scattering" domain in heavy ion collisions. The data on another
penetrating probe, direct photons, still remain the upper bound.

\section{J/$\psi$ suppression}
 Very impressive PbPb data from NA50 have been presented here \cite{Fleuret},
 and then discussed extensively by many theorists. 

Basically there are two alternatives: (i) the ``early scenario'' 
\cite{Blaizot} which
blames destruction of J/$\psi$ on QGP; and (ii) the ``late scenario''
\cite{late}
which
relates the difference between S and Pb data on much larger number of
final
state collisions in the latter case.

 $If$  the J/$\psi$
absorption cross section on mesons is of the order of 1-3 mb, it was
demonstrated \cite{late} that one can fit the data.
But is this number  really true? It was noticed long ago that the
world
of heavy and light quark hadrons does not speak much to each
other\footnote{
For those who want an impressive experimental example: the rate of
$\psi'\rightarrow\psi\pi\pi$ decay is about thousand times less than
$\rho'\rightarrow\rho\pi\pi$.}. Theoretical explanation of this fact
is that the former exist due to confinement and Coulomb-type color
exchange, while the latter are mostly bound by forces responsible for
chiral breaking
(instantons\footnote{Although
the latter effects are 
 stronger and much better understood by theorists, they are also
much less known.}). 

   Two experiments can significantly clarify the situation: (i) To
   run PbPb collisions at somewhat lower energy. If the early
scenario is right and
   the energy behavior has a sharp threshold, one should be able to
   see it. (ii) To perform the inverse kinematics
experiment, Pb on p, and see what happens with $J/\psi$ at rest in
matter.

   The issue of   $J/\psi$ $p_t$ distribution was discussed by
    R.Vogt. Her calculations are based on the initial-state
   re-scattering of an incoming gluon in nuclei. The prediction is that
   additional transverse momentum is\footnote{Experimentally the
     constant $\lambda^2_{\psi}$ is smaller than
     $\lambda^2_{\Upsilon}$:
why this is so was discussed in \cite{Peigne}. } 
 $<p_t^2>-<p_t^2>_{NN}=\lambda^2_{\psi} (\bar n -1)$
 , where $\bar n$ is the total number of nucleons on the way. 
It works for pA and SA reactions: if it would be confirmed in PbPb
as well, it would indeed mean that at least $elastic$ scattering of   $J/\psi$
on
mesons is small (with obvious limits on absorption from optical theorem).
The opposite alternative is strong hadronic rescattering: in this case
 $J/\psi$ (or some fraction of them) would tend to be ``thermalized'',
 as other hadrons.

  The final point I would like to make deals with  $\psi'/ J/\psi$
  ratio. Centrality dependence in PbPb of this ratio suggests that it
is no more  
decreasing. Can it be that
 we have hit the bottom, all initial   $\psi'$ are suppressed
and  all the observed   $\psi'$ are actually
coming from an excitation of the  $J/\psi$ itself? 
See discussion of how this excitation may happen in \cite{SSZ}.

\section{Hadronic observables}

  For long time the major question was whether heavy ion collisions really
produce a ``hadronic matter", a reasonably well thermalized one. 
Studies of particle composition have indeed shown that (apart of rare species like
multistrange ones) it agrees with the thermal model well enough. The extracted
``chemical freeze-out" parameters at AGS and SPS
\cite{Stachel} both happen to be very close to the expected
line of the QCD phase transition\footnote{This agrees well with the theoretical ideas suggesting ``dropping masses" and large cross sections, at least for $\sigma$ exchanges. }

  The next logical question is whether this matter is in a state of 
``collective motion", so that (at least at sufficient late stages) one can use
simple hydrodynamical language to describe it.
With the arrival of Au-beam data at BNL and Pb-beam
 ones at CERN, we can definitely answer ``yes" to this question.
In fact, hydro motion is so complicated that it became necessary to divide
it into 4 categories now:
(i) the
longitudinal motion; (ii) radial (axially symmetric) one; (iii) dipole and
(iv) quadrupole (or ``elliptic") one in polar angle.

  The  longitudinal motion is  usually related to (somewhat
vague) 
issue known as
``stopping". One way to address it is to ask whether matter
elements\footnote{It is implied that the thermal motion (depending on
the particle mass and thus different for different secondaries) is
deconvoluted from the distribution.}
with different rapidities have all the same or different composition.
For light ions the answer  definitely is negative:  near the beam rapidity
the
baryon/meson ratio is much larger than at midrapidity. For
heavy ion data this ratio is rather constant. I think the question is not
yet
quite answered though,
and better data on other secondaries such as K,d etc
are needed to decide how collective  the longitudinal motion really is.

In contrast to the longitudinal flow, a  $radial$ one is very well
documented. The transverse mass slope is measured for
$\pi^+,\pi^0,\pi^-,K^+,K^0,K^-,p,\bar p, \Lambda, d$, it is independent on
particle charge but consistently increasing with the particle mass,
indicated a common flow\footnote{Studies of the question inside the event
generators such as RQMD lead to the same conclusion.}. The effect is
increasing strongly with A, and is confined to central rapidities.
There seem to be little difference between the radial flow at  AGS and SPS.

   Quantitative study of all these data are only started, but it is
already clear that they allow to
 address the fundamental
  question 
  about the Equation of State
(EOS) of hot/dense hadronic matter.
 Significant part of my original talk is dealing with new development
along this line,
 a  model for AGS/SPS energy domain, called {\it Hydro-Kinetic
   Model}, HKM. The main conclusion coming from this (still unfinished work)
is that at AGS the ``standard" EOS (the resonance gas plus the QGP) is too 
soft, while for SPS data (especially the NA44 and NA49 ones on the slopes of $m_t$ distributions) it actually does a good job\footnote{  
Cascade event generators 
(such as Venus,RQMD,ARC
  etc) also have their own  EOS as well. For example,
 Venus 
and RQMD  also are rather ``soft" at the early stages: although  
they have no QCD phase transition,  their longitudinally stretched strings
 do not lead to
the transverse pressure. 
}. The no-phase-transition scenario
with the resonance gas only produce too much flow, and therefore should be
considered experimentally excluded.

  A dipole component of the flow (in the collision plane) was found at
AGS but not so far at SPS: however the quadrupole one was recently detected at
both
energies. As it was originally emphasized in \cite{Olli}, the latter one is
especially
sensitive to EOS, and its further studies are clearly of great interest. 

   The latest stage of the collision is known as a ``thermal"
freeze-out: after that even elastic re-scattering of secondaries are frozen.
Although this issue is related with ``known physics" at the most dilute stage,
this issue continue to be  a matter of on-going debates.
In particular, Gaardhoje (NA44) has 
argued that their data are best fitted with some universal freeze-out temperature $T_f=140 MeV$, which would mean that the interaction stops very soon after
passing the phase transition line into the hadronic phase. However, a number of arguments suggest that in PbPb collisions the system actually is large enough
to cool further, to  $T_f=110-120 MeV$.

  One such indication in fact was included in  Gaardhoje's talk: it is recent analysis \cite{Coul} of the Coulomb effects  seen in $\pi^+/\pi^-$
ratio at small $p_t$. The second one follows from HBT radii (see talks by Foka and Wiedemann). Two more arguments were presented in my talk: one is direct
application of kinetics of the freeze-out, another is related with
a very strong A-dependence of radial flow. I have argued that ``extra push"
in PbPb case may only come  from the latest  hadronic stage
of the evolution.

\section{Jet energy losses in matter and Landau-Pomeranchuck-Migdal effect}

   These issues were the subject of several theoretical talks. Shulga reviewed
the situation in QED,  pointing out that recent SLAC experiments have confirmed
not only the original LPM predictions but also a thin-target predictions made
by him and collaborators. 

  Baier has described some theoretical problems with ``hard loop re-summation",
which appears in higher order processes in QGP with photon and dilepton emission. He then go on to describe what is known as BDMPS mechanism
\cite{BDMPS} of the energy loss calculation, the QCD analog of LPM effect. He has explained that there is a very significant difference between QCD and QED: it is gluons which mostly experience rescatterings and therefore a color charge is constantly ``undressing"
in matter. As a result, the conventional idea of dE/dx does not hold and the
energy loss is proportional to the $second$ power
of distance L traveled in matter: for estimates one can use
$$ -\Delta E_g=(10-20) (L/10 fm)^2$$
 B.Zakharov (using a different formalism)
has studied how the effective power of L changes from 2 to 1 for thin targets.

  This large energy loss of jets is very important because it supports the idea
of very rapid thermalization of QGP at RHIC/LHC. It can be directly
  observed by looking at jet shape (see talk by S.Gupta). Another important consequence is that charmed quarks emitted early in hard processes got ``stuck"
in matter:  this basically kills the correlated
charm background to  the dilepton signal at RHIC \cite{Shu_charm}.



\end{document}